# Ultra-Sensitive Hot-Electron Nanobolometers

# for Terahertz Astrophysics

Jian Wei [a,1], David Olaya [a,1], Boris S. Karasik [b,1], Sergey V. Pereverzev [a], Andrei V. Sergeev [c], and Michael E. Gershenson [a]

[a] Department of Physics and Astronomy, Rutgers University,

136 Frelinghuysen Rd., Piscataway, NJ 08854, USA

[b] Jet Propulsion Laboratory, California Institute of Technology, Pasadena, CA 91109, USA

[c] SUNY at Buffalo, Buffalo, NY 14260, USA

**The background-limited spectral imaging of the early Universe requires spaceborne terahertz (THz) detectors with the sensitivity 2-3 orders of magnitude better than that of the state-of-the-art bolometers[1,2]. To realize this sensitivity without sacrificing operating speed, novel detector designs should combine an ultrasmall heat capacity of a sensor with its unique thermal isolation. Quantum effects in thermal transport at nanoscale [3] put strong limitations on the further improvement of traditional membrane-supported bolometers [4]. Here we demonstrate an innovative approach by developing superconducting hot-electron nanobolometers in which the electrons are cooled only due to a weak electron-phonon interaction. At $T<0.1$K, the electron-phonon thermal conductance in these nanodevices becomes less than one percent of the quantum of thermal conductance. The**

---

[1] J.W., D.O. and B.S.K. contributed equally to this work.

**hot-electron nanobolometers, sufficiently sensitive for registering single THz photons, are very promising for submillimeter astronomy and other applications based on quantum calorimetry and photon counting.**

Since the original invention by Samuel P. Langley in 1878 [5], bolometers have gone a long way of improving the sensitivity and expanding the frequency range, from X-rays and optical/UV radiation to the submillimeter waves. The latter range contains approximately half the total luminosity of the Universe and 98% of all the photons emitted since the Big Bang [6]. Because the performance of ground-based THz telescopes is severely limited by a strong absorption of THz radiation in the Earth atmosphere, the development of space-based THz telescopes will be crucial for better understanding of the Universe evolution. Active cooling of primary mirrors on these telescopes will reduce the mirror emission below the cosmic background level (Fig. 1) and greatly expand the range of observable faint objects. The development of advanced detectors with background-limited sensitivity for such telescopes poses a significant challenge. Indeed, the photon flux $N_{ph}$, which corresponds to the cosmic background fluctuations, is very weak: at $\nu > 1$ THz, the photon flux in a diffraction-limited beam does not exceed 100 photons/s for typical extragalactic emission lines with $\nu/\delta\nu \sim 1000$. The noise equivalent power (*NEP*) of a background-limited detector should be less than $NEP_{ph} = h\nu\sqrt{2N_{ph}} \sim 10^{-20}$ W/Hz$^{1/2}$, which is a factor of 100 lower than that of state-of-the-art bolometers.

Although new detector concepts are coming into play nowadays [7,8], bolometers still have a great potential for achieving the most challenging goals. Realization of the ultra-high sensitivity requires an unprecedented thermal isolation of a bolometer. Indeed, in the flux-integrating regime (the bolometric time constant $\tau \gg N_{ph}^{-1}$), the minimum *NEP* is determined by the thermal energy fluctuations in a bolometer, and the corresponding value of $NEP_{TEF} = \sqrt{4k_B T^2 G}$ is controlled by the thermal conductance $G$ between the bolometer and its environment. In a traditional (the so-called "geometrically isolated") bolometer, $G$ is determined by the number of relevant phonon and photon "channels" (modes) participating in thermal transport between the sensor and its environment. It has been shown recently for both photons [3,9] and phonons [10] that the thermal conductance of a short single channel is determined by the universal value $G_Q = \pi^2 k_B^2 T/3h$ known as the quantum of thermal conductance. However, even a single-mode coupling is too strong for the realization of the required bolometer sensitivity in the temperature range accessible for practical space instruments (down to 50 mK). The reduction of the photon conductance can be achieved by enhancing an electrical impedance mismatch between the bolometer and its surroundings at frequencies $f \sim k_B T/h$. The phonon conductance can be decreased by increasing the channel length. The resources of the latter approach are rather limited, as recent experiments have shown: the minimum value $G \approx 0.2 G_Q$ has been observed at 0.09K for a bolometer supported by very long (~ 1 cm) and narrow (~ 1 μm) $Si_3N_4$ beams. Despite a relatively small size of this micromachined device, the heat capacity $C$ was still rather large, which resulted in a slow bolometric response with the time constant $\tau = C/G = 1\text{-}10$ s.

Here we present a novel approach that enables a significant increase of the bolometer sensitivity and, at the same time, reduction of its response time. Fast response in a well isolated bolometer requires a very small heat capacity $C$ and, thus, the *nanoscale* dimensions of a sensor. To overcome the limitation of fast phonon exchange, we realized the hot-electron regime [11,12,13] in superconducting nanobolometers at sub-Kelvin temperatures. In this case, a weak electron-phonon coupling, which governs the effective thermal conductance, dramatically improves the thermal isolation of the sensor [14]. This approach allowed us to simultaneously realize

$G \sim 0.01 G_Q$ and $\tau \sim 0.3$ ms at $T = 0.06$ K. Moreover, an ultra-small electron heat capacity of these nanobolometers ($C \sim 10^{-19}$ J/K at $T = 0.1$ K) makes them suitable for registration of single THz photons.

We fabricated the nanobolometers using electron-beam lithography and e-gun deposition of Ti and Nb films on Si substrates covered with a layer of $SiO_2$ or $Si_3O_4$ (for details, see Supplementary Information). The device consists of a titanium "island" with a volume of $\sim 10^{-2}$ μm$^3$ flanked by niobium current leads (see the inset in Fig. 2). *In-situ* deposition of Ti and Nb at different angles through a "shadow" mask ensured the low resistance of the Ti/Nb interface. The critical temperature $T_C(Ti) = 0.1$-$0.2$ K for thin ($d \approx 40$ nm) and narrow ($w \approx 0.1$ μm) Ti nanobridges was lower than the "bulk" value (0.4 K). The parameters of several devices are listed in Table 1.

These devices operate at sub-Kelvin temperatures in the hot-electron regime at the temperature $T < T_C(Ti)$. Absorption of radiation by electrons in the titanium "island" leads to an increase of their effective temperature $T_e$ and to a corresponding change in resistance while the crystal lattice remains in equilibrium with the bath. Superconducting leads with high critical temperature dramatically improve thermal isolation of electrons in the nanosensor: the outdiffusion of hot electrons with energies $\varepsilon < \Delta_{Nb} \sim 1.3$ meV ($\Delta_{Nb}$ is the superconducting gap in Nb) from the nanobridge into the current leads is blocked by Andreev reflection at the Ti/Nb interface [15].

To characterize the device performance, we studied the thermal conductance, $G$, the response time, $\tau$, and the electron heat capacity, $C_e$. The thermal conductance $G(T)$ was found by measuring an increase of the electron temperature $T_e$ induced by dc current (for details, see Supplementary Information). Though this method of measuring $G$ is well known, its application to the present devices required very low levels of heating power (even $P \sim 10^{-18}$ W was sufficient to overheat some devices above $T_C$). Application of this method at the lowest temperatures also required thorough filtering of the external noise over a wide frequency range; the estimated noise power released in our samples did not exceed $10^{-19}$ W.

The thermal conductance $G(T)$ normalized by the volume of Ti nanosensors, $V$, is shown for several devices in Fig. 2. Both the temperature dependence of $G(T)$ and the $G/V$ values agree very well with the measurements of the thermal conductance in much larger ($\sim 10^4 \times 2 \times 0.04$ μm$^3$) meander-line Ti devices, where $G$ is solely determined by the electron-phonon coupling [16]. In disordered ultra-thin films, this coupling originates from the electron scattering by vibrating defects and boundaries and, therefore, is sensitive to the film morphology and substrate material. Depending on the nature of disorder, the low temperature thermal conductance varies between $G_{min} \sim T^5$ and $G_{max} \sim T^3$ [17]; modification of phonon spectra in ultrathin films also affects the relaxation [18].

For the devices studied in this work, the dependence $G(T)$ was approximately proportional to $T^3$. Note that $G$ close to $G_{min}$ has been previously observed for large Ti devices on sapphire substrates ; this indicates that $G$ in optimized nanodevices can be further reduced.

The observed scaling of $G$ with the sensor volume over 6 orders of magnitude provides an experimental proof that (a) the phonon emission governs the electron energy relaxation in the studied nanosensors, and (b) the diffusion of hot electrons is effectively blocked by Andreev reflection. The lowest thermal conductance $G \sim 10^{-16}$ W/K, which we observed at 40 mK, is $\sim 400$ times smaller than $G_Q$. In particular, observation of such an ultra-low thermal conductance suggests that the emission of photons by hot electrons [3,9] was drastically reduced in our devices due to a large electrical impedance mismatch between the nanobolometer and its environment at frequencies $f \sim$ 1-10 GHz which correspond to black body radiation at $T$ = 0.1-0.3 K.

In order to disentangle all three quantities in the expression $G = C_e/\tau$, we performed an independent measurement of the response time using a dc SQUID amplifier with a 120 kHz bandwidth. Figure 3 shows the response of a device to an absorption of a single photon with $\lambda$ = 1.55 μm. For these measurements, we used devices with a large volume ($\sim$ 1.5 μm$^3$) to ensure that the increase of the electron temperature $\Delta T_e = h\nu/C_e$ is always smaller than the width of the superconducting transition, $\delta T_C$. Good agreement was observed between the experimental values of $\tau$ and the ratio $C_e/G$ calculated using the experimental values of $G(T)$ and the electron heat capacity $C_e = \gamma TV$ with $\gamma$ = 310 W·m$^3$·K$^{-1}$ (the "bulk" value for Ti).

The unparalleled thermal isolation of electrons leads to a very low intrinsic detector noise in both the flux-integrating and photon-counting regimes. For example, the condition for the flux-integrating operation ($N_{ph}\tau \gg 1$) is met for device 1 at $\nu \leq$ 600 GHz (see Table 1 and Fig. 1). The value of $NEP_{TEF}$ for this device, $9 \times 10^{-21}$ W/Hz$^{1/2}$ at $\sim$ 65 mK, is well below the limits imposed by background fluctuations.

Above 1 THz, because of a very low cosmic background (Fig. 1), the background-limited sensitivity of nanobolometers can be realized only in the photon-counting regime [19] ($N_{ph}\tau \ll 1$). An ultra-small heat capacity $C_e$ is the key factor for the realization of the minimum detectable energy $\delta\varepsilon \approx \sqrt{k_B T_C^2 C_e}$ sufficiently small for the registration of single THz photons (see Table 1). Indeed, the dark counts in hot-electron bolometers are caused by the fluctuations of electron energy; these fluctuations with correlation time $\tau$ follow the Gaussian distribution with width $\delta\varepsilon$. Two basic parameters of a counter, the dark count rate, $N_d$, and the intrinsic quantum efficiency,

$\eta$, are interrelated: the trade-off between these parameters is set by the threshold energy $E_T$ of the readout process (see Supplementary Information). If $h\nu/\delta\varepsilon \gg 1$, both high $\eta$ and low $N_d$ can be simultaneously realized by setting $E_T$ a few times greater than $\delta\varepsilon$. If $\delta\varepsilon \sim h\nu$, a careful adjustment of $E_T$ may be required to optimize $N_d$ and $\eta$.

Table 1 illustrates the performance of our devices in the photon-counting mode for $\nu = 1.9$ THz (spectral line corresponding to the $^2P_{3/2} \rightarrow {}^2P_{1/2}$ atomic fine-structure transition of ionized carbon) with spectral resolution $\nu/\delta\nu \sim 1000$. The last column in Table 1 presents the values of $\eta$ corresponding to the background limited detection. For all 4 devices with the critical temperature values not exceeding 200 mK the quantum efficiency is close to 100%. For higher critical temperature or lower radiation frequency the quantum efficiency may become lower due to a smaller value of $h\nu/\delta\varepsilon$ ratio. Strong dependence of $\eta$ on $E_T/\delta\varepsilon$ provides an interesting possibility for making this detector adaptive to the photon flux. This would be important in large imaging arrays where the luminosity may substantially vary from pixel to pixel.

Overall, the achieved device characteristics demonstrate a strong potential of the hot-electron nanobolometer for the most challenging THz astronomical applications. Relatively large surface impedance of thin Ti films makes the integration of nanobolometers with THz microantennas rather straightforward [20]. The SQUID-based multiplexed readouts, which are becoming readily available for large-format superconducting imaging arrays [21], are also suitable for the hot-electron bolometers. Besides astrophysics, these novel nanodevices are promising for other THz applications that require photon (and phonon) counting and quantum calorimetry [22].

**Acknowledgements**

We thank J.H. Kawamura at JPL/Caltech for help with experiments. The work at Rutgers was supported in part by the NASA grant NNG04GD55G, the Rutgers Academic Excellence Fund, and the NSF grants. The research at the Jet Propulsion Laboratory, California Institute of Technology, was carried out under a contact with the National Aeronautics and Space Administration.

**Table 1. Parameters of the hot-electron nanobolometers.**

| Device | Dimensions length × width × thickness, μm³ | $T_C$ mK | $\delta T_C$ mK | $G(T_C)$ $10^{-15}$ W/K | $C_e(T_C)$ $10^{-20}$ J/K | $\tau(T_C)$ μs | $\delta\varepsilon/h$ THz | $NEP_{TEF}$ $10^{-21}$ W/Hz$^{1/2}$ | $\eta$ @1.9THz |
|---|---|---|---|---|---|---|---|---|---|
| 1 | 1.0 × 0.13 × 0.04 | 65 | 5 | 0.33 | 9 | 320 | 0.12 | 9 | 1.0 |
| 2 | 0.64 × 0.11 × 0.04 | 130 | 50 | 5.4 | 11 | 20 | 0.25 | 70 | 1.0 |
| 3 | 1.0 × 0.11 × 0.04 | 200 | 50 | 24 | 25 | 10 | 0.56 | 230 | 0.8 |
| 4 | 4.0 × 0.10 × 0.04 | 130 | 30 | 13 | 64 | 50 | 0.59 | 110 | 0.9 |

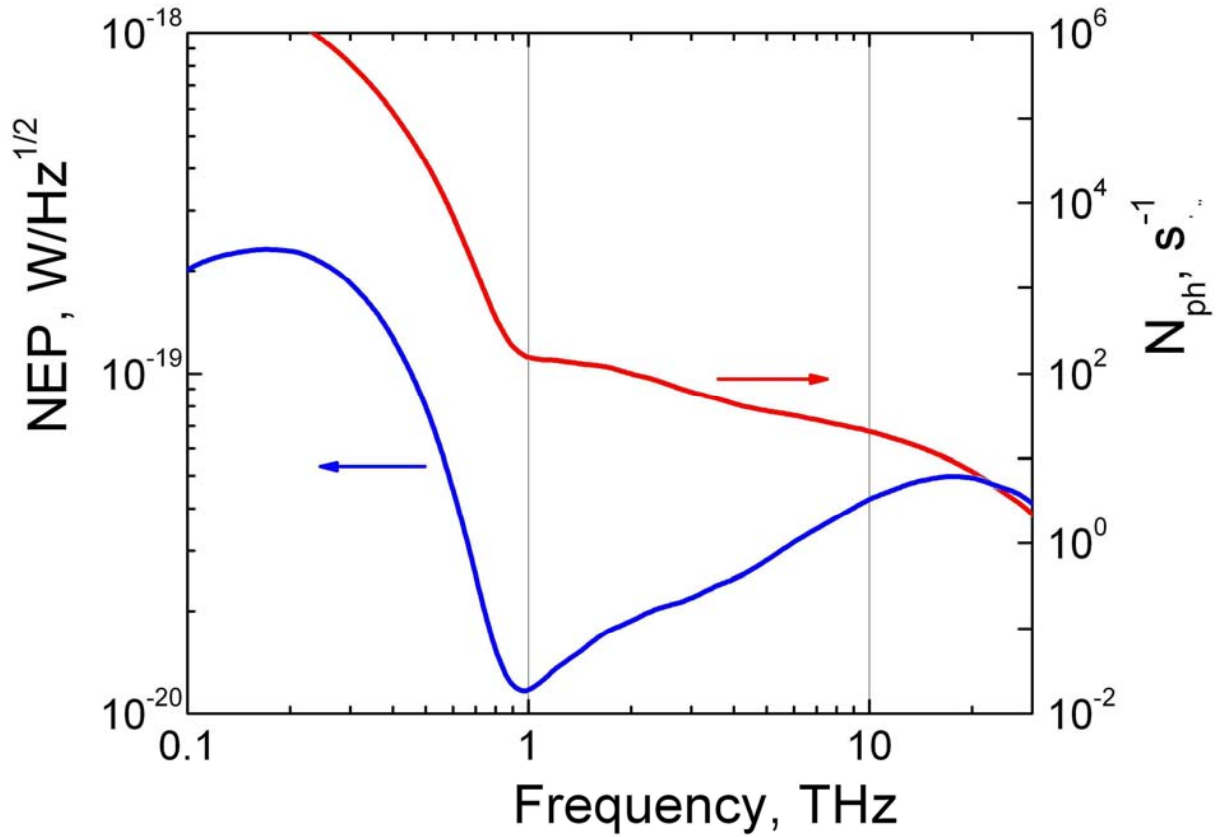

**Fig. 1. The background–limited detector sensitivity for THz spectroscopy in space.**
The background photon flux $N_{ph}$ (red curve) is shown for spectroscopic observations within a diffraction-limited beam with resolution $\nu/\delta\nu \sim 1000$, which corresponds to the width of a typical extragalactic emission line. The background flux is very weak, $N_{ph} < 100$ photons/s above 1 THz. The background-limited *NEP* (blue curve) has been calculated from the experimental background luminosity [23] assuming single polarization and an optical coupling efficiency of 25%. The detector performance is limited by the Cosmic Microwave Background at $\nu < 1$ THz, and the radiation from the galactic core and dust clouds at higher frequencies.

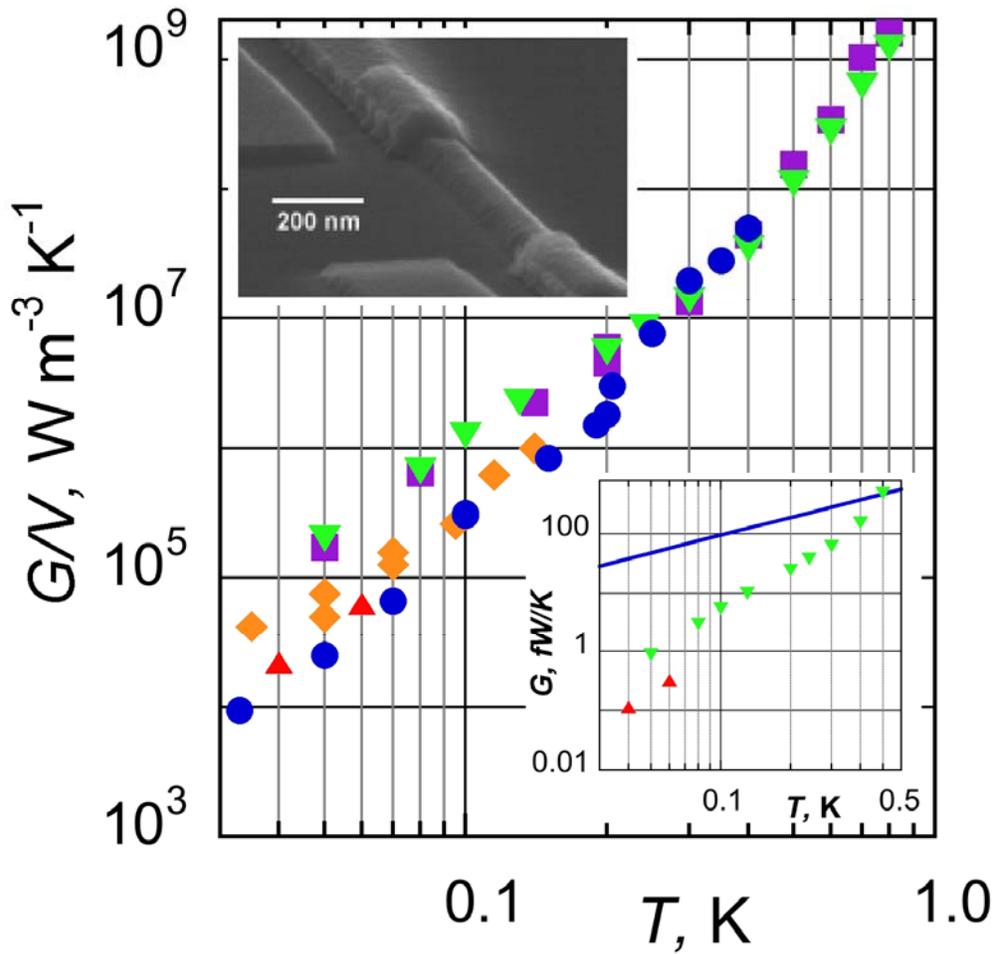

**Fig. 2. The thermal conductance for hot-electron nanobolometers.**

The normalized-by-volume thermal conductance was measured for nanobolometers 1 (▲), 2 (▼), 3 (■), 4 (♦), and for a large meander-line device (●) over the temperature range $T = 0.03$-1K (for the device parameters, see Table 1). The top inset shows the SEM image of a Ti/Nb nanosensor with dimensions $0.04\mu m \times 0.14\mu m \times 0.56\mu m$. The Ti nanobridge is flanked by Nb current leads, the rest of Ti film is separated from Ti nanosensor and Nb leads by trenches. The bottom insert shows the absolute values of $G$ for nanobolometers 1 and 2, together with the quantum of thermal conductance $G_Q = \pi^2 k_B^2 T / 3h$ (blue line).

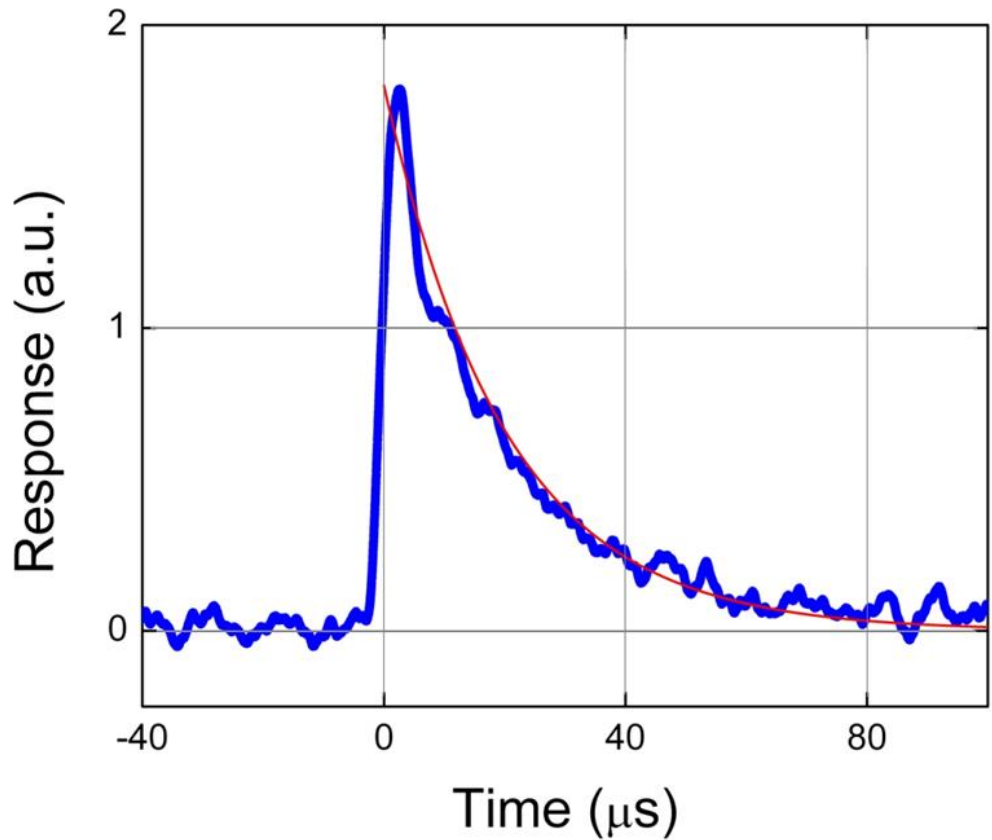

**Fig. 3**. **Detection of a single near-IR photon by a hot-electron bolometer**.

The response of a hot-electron bolometer with dimensions $20 \times 0.10 \times 0.04$ μm$^3$ to an absorption of a single photon of radiation with $\lambda = 1.55$ μm. At $T = 0.19$ K, the relaxation time $\tau$ is approximately 20 μs, which is close to the electron cooling time due to electron-phonon coupling. Red line is the exponential fit of the signal decay.

**Supplementary Information**

**Device Fabrication.** The nanostructures were fabricated on Si substrates covered by a ~ 0.2μm-thick layer of $SiO_2$ or $Si_3N_4$. Titanium and niobium films were deposited by an electron-gun source in an oil-free vacuum system with base pressure ~$1\times10^{-9}$ mbar. For fabrication of an oxide-free Ti/Nb interface, we have used the so-called "shadow mask" technique: Ti and Nb films were sequentially deposited at different angles through a "shadow" mask without breaking vacuum. An SEM image of a "shadow" mask prepared by electron-beam lithography is shown in Fig. 1A. To reduce substrate/resist heating in the process of deposition of Nb (melting point $T = 2468°C$), the distance between the substrate holder and the e-gun source was increased up to 40 cm. The rotary substrate holder can position the substrate at any angle with respect to the direction to the e-gun source. Thin Ti film was deposited at a rate ~1.5-3 nm/s along the direction normal to the plane of the substrate, Nb was deposited at a rate 0.5 nm/s along the direction which was at a 45° angle to the substrate and perpendicular to the long dimension of the nanosensor. Because the resist thickness exceeds the width of the narrow channel, Nb film covers only the ends of Ti nanosensor.

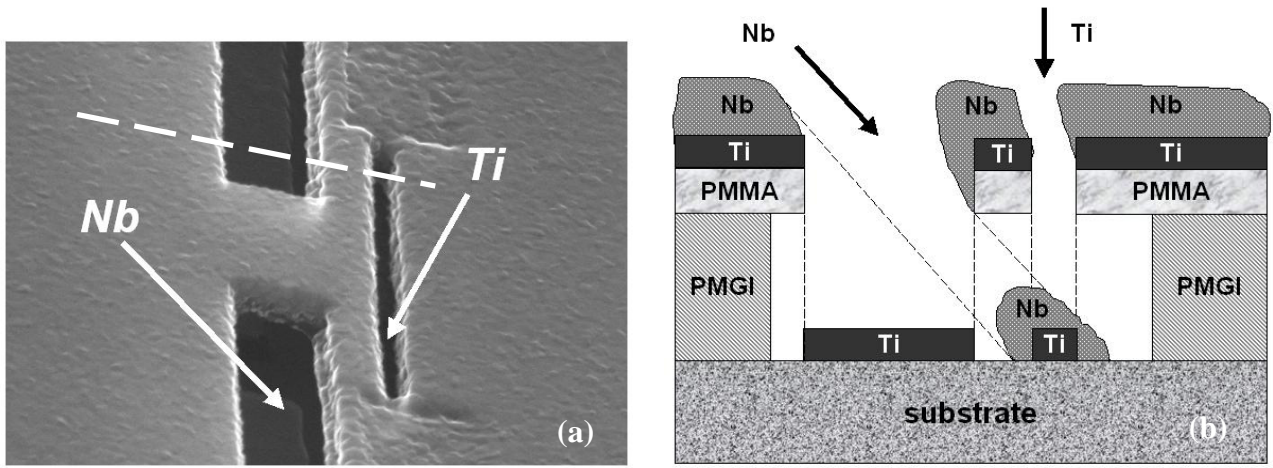

**Supplementary Figure 1**. **(a) Shadow mask used for the fabrication of Ti nanobolometers.** The SEM picture has been taken after the deposition of Ti and Nb films. The bi-layer lift-off mask consists of the top 0.1-µm-thick PMMA layer, used as a high-resolution electron resist, and the bottom 0.3-µm-thick layer of polymer PMGI. A deep undercut in the bottom layer allows for overlapping the films deposited at different incident angles. **(b) Cross section along the dashed line on panel (a)**. Titanium was deposited along the direction normal to the plane of the substrate, Niobium was deposited at 45° incidence angle and perpendicular to the long dimension of the 1-µm-long slit.

**Measurements of the thermal conductance.** To measure the thermal conductance $G(T)$, the Ti nanosensor was overheated by a small dc current $I_{dc}$, and the corresponding difference between the electron temperature $T_e$ and the equilibrium bath temperature $T_{ph}$ was recorded; $G(T)$ was estimated from the heat balance equation, $RI_{dc}^2 = G(T_e)\Delta T_e$ ($T_e - T_{ph} \ll T_{ph}$). The change in $T_e$ was determined from the heating-induced change of the resistance using the dependence $R(T)$ measured in equilibrium as a resistance thermometer. The resistance of devices was measured by an ac (13 Hz) resistance bridge using a small (typically, 0.1-1 nA) measuring current. Above $T_C$, this dependence $R(T)$ is due to the proximity effect at Ti/Nb interfaces, below $T_C$ it is due to the resistive state of a superconductor induced by the magnetic field perpendicular to the plane of Ti nanobridge. In the measurements at $T < T_C$, a fixed value of $R \sim 0.9\, R_N$ ($R_N$ is the resistance in the normal state) was maintained by applying a magnetic field close to the critical field at a given temperature. This method assumes that the non-equilibrium electron distribution function can be characterized with an effective electron temperature. This assumption is valid at sub-Kelvin temperatures, where the electron-electron scattering rate in thin films exceeds by many orders of magnitude the electron-phonon scattering rate.

**Evaluation of the photon counter performance.** Thermodynamic fluctuations of the electron energy in hot-electron bolometers are characterized by the Gaussian distribution with width $\delta\varepsilon \approx \sqrt{k_B T_C^2 C_e}$. Absorption of a photon "shifts" this distribution by $h\nu$ toward higher energies. The probability to find the electron system in a state with energy greater than the discrimination threshold energy $E_T$, which is determined by the detector bias and/or the readout electronics, is given by the Gaussian integral:

$$\eta = \int_{(E_T - h\nu)/\delta E}^{\infty} \frac{dx}{\sqrt{2\pi}} \exp\left(-x^2/2\right). \tag{1A}$$

Equation 1A shows that the quantum efficiency of the bolometer, $\eta$, depends on two dimensionless parameters, $E_T/\delta\varepsilon$ and $h\nu/\delta\varepsilon$. In its turn, the number of dark counts, $N_d$, which are caused by the energy fluctuations with an amplitude greater than $E_T$, is given by

$$N_d = B \int_{E_T/\delta E}^{\infty} \frac{dx}{\sqrt{2\pi}} \exp\left(-x^2/2\right), \tag{2A}$$

where $B \sim \tau^{-1}$ is the detection bandwidth.

For $h\nu/\delta\varepsilon \gg 1$, by choosing $E_T$ a few times greater than $\delta\varepsilon$, one can substantially reduce the dark count rate and, at the same time, realize the quantum efficiency close to 1. For $\delta\varepsilon \sim h\nu$, the choice of $E_T$ is dictated by the trade-off between $\eta$ and $N_d$.

In order to realize the background-limited sensitivity, the average fluctuation of the number of intrinsic dark counts, $\sqrt{N_d t}$, during the observation time $t$ should be smaller than the fluctuation of the number of counts due to the background photon flux, $\eta\sqrt{N_{ph} t}$. Thus, the following condition should be fulfilled:

$$N_d / \eta^2 \leq N_{ph}. \qquad (3A)$$

Equations 1A-3A have been used to evaluate the performance of the studied nanobolometers (see Table 1).